\documentclass[useAMS,usenatbib]{mn2e}
\usepackage{times}
\usepackage{graphicx}

\usepackage{amsmath}
\usepackage{amssymb}

\usepackage[T1]{fontenc}
\usepackage{aecompl}
\usepackage{epstopdf}

\begin{document}

\title[Massive Pop III Galaxies via Photoionization]{Formation of Massive Population III Galaxies through Photoionization Feedback: A Possible Explanation for CR7}

\author[E. Visbal et al.]{Eli Visbal$^1$\thanks{visbal@astro.columbia.edu} \thanks{Columbia Prize Postdoctoral Fellow in the Natural Sciences}, Zolt\'{a}n Haiman$^1$, Greg L. Bryan$^1$ \\ $^1$Department of Astronomy, Columbia University, 550 West 120th Street, New York, NY, 10027, U.S.A. }

\maketitle

\begin{abstract}
We explore the formation of massive high-redshift Population III (Pop III) galaxies through photoionization feedback. We consider dark matter halos formed from progenitors that have undergone no star formation as a result of early reionization and photoevaporation caused by a nearby galaxy. Once such a halo reaches $\approx 10^9~M_\odot$, corresponding to the Jeans mass of the photoheated intergalactic medium (IGM) at $z\approx 7$, pristine gas is able to collapse into the halo, potentially producing a massive Pop III starburst. We suggest that this scenario may explain the recent observation of strong He II 1640~\AA~line emission in CR7, which is consistent with $\sim 10^7~M_\odot$ of young Pop III stars. Such a large mass of Pop III stars is unlikely without the photoionization feedback scenario, because star formation is expected to inject metals into halos above the atomic cooling threshold ($\sim 10^8~M_\odot$ at $z \approx 7$). We use merger trees to analytically estimate the abundance of observable Pop III galaxies formed through this channel, and find a number density of $\approx 10^{-7}~{\rm Mpc^{-3}}$ at $z=6.6$ (the redshift of CR7). This is approximately a factor of ten lower than the density of Ly$\alpha$ emitters as bright as CR7. 
\end{abstract}

\begin{keywords}
stars:Population III--galaxies:high-redshift--cosmology:theory
\end{keywords}

\section{Introduction}
In the context of the standard model of cosmology, a theoretical picture for cosmic dawn has started to emerge. Numerical simulations predict that the first Pop III stars form in $\sim 10^5~M_\odot$ ``minihalos'' through molecular hydrogen cooling \citep[for a recent review see][]{2015ComAC...2....3G}. As the global star formation density increases, a background of Lyman-Werner (LW) radiation builds up over time. Once it reaches sufficient intensity, this background photodissociates molecular hydrogen, suppressing subsequent Pop III star formation in the smaller minihalos \citep[e.g.][]{1997ApJ...476..458H, 2001ApJ...548..509M, 2007ApJ...671.1559W, 2008ApJ...673...14O, 2011MNRAS.418..838W, 2014MNRAS.445..107V}. Eventually the LW flux is strong enough to prevent star formation in all minihalos. At this point, Pop III stars are only expected to form in ``atomic cooling'' halos with $T_{\rm vir} \gtrsim 10^4~{\rm K}$ (corresponding to $\sim 10^{8}~M_\odot$ at $z=7$), where cooling can proceed via atomic hydrogen transitions. Generally it has been considered unlikely for Pop III stars to form in halos much larger than the atomic cooling threshold because rapid self-enrichment from metals quickly leads to Pop II star formation \citep[e.g.][]{2012ApJ...745...50W}. 

To date there have been no unambiguous direct detections of Pop III stars (though see \cite{2015MNRAS.453.4456V} for limits based on \emph{Planck} obserations). This may have been changed by the discovery of CR7, a bright Ly$\alpha$ emitter at $z=6.6$ \citep{2015ApJ...808..139S}. CR7 has strong Ly$\alpha$ and He II 1640~\AA~emission lines, and no detected metal lines, as predicted for Pop III stars \citep[see e.g.][]{2001ApJ...553...73O, 2001ApJ...550L...1T}. It consists of three distinct clumps (A, B, and C) separated by a projected distance of $\sim 5~{\rm kpc}$.  \cite{2015ApJ...808..139S} find that the observation is consistent with $\sim10^7 ~M_\odot$ of Pop III stars in clump A (though this mass can vary depending on the initial mass function) and $\sim10^{10} ~M_\odot$ of older metal-enriched ($0.2~Z_\odot$) stars in clumps B and C.  The presence of such a large cluster of Pop III stars is puzzling since, even assuming saturated LW feedback, star formation is expected to start at the atomic cooling threshold, promptly leading to self-pollution and Pop II star formation. As an alternative interpretation, several groups have suggested that clump A consists of a so-called direct collapse black hole (DCBH), accreting pristine gas \citep{2015MNRAS.453.2465P, 2015arXiv151001733A, 2015arXiv151201111H}. Note that clump A is spatially resolved by Hubble Space Telescope observations. The fact that it is spatially extended may favor a stellar interpretation.

In this letter, we explore a formation scenario for massive Pop III clusters which can potentially explain how CR7 could contain such a high mass of Pop III stars. We consider a $\sim 10^9~M_\odot$ dark matter halo which forms in an environment where the IGM is reionized well before the formation of the halo. In particular, we consider a region where reionization occurs early enough and with sufficient ionizing flux such that the gas in the progenitors of the $\sim 10^9~M_\odot$ halo is photoevaporated before any stars can form. This ionized gas is photoheated to $\sim 10^4~{\rm K}$ leading to a Jeans mass of $M_{\rm J} \gtrsim 10^9 M_\odot$. This prevents gas from collapsing into the halo until it reaches $M_{\rm J}$, at which point it falls back in, potentially leading to a rapid burst of Pop III star formation. We perform a rough estimate of the number density of Pop III galaxies which can form this way and determine that it may be sufficient to explain the inferred number density of objects similar to CR7. For a related study on Pop III galaxies formed through reionization see \cite{2010MNRAS.404.1425J}. Throughout we assume a $\Lambda$CDM cosmology consistent with constraints from \emph{Planck} \citep{2014A&A...571A..16P}: $\Omega_\Lambda=0.68$, $\Omega_{\rm m}=0.32$, $\Omega_{\rm b}=0.049$, $h=0.67$, $\sigma_8=0.83$, and $n_{\rm s} = 0.96$. When quoting space number densities we use comoving units, and for distances we use physical units.

\section{Massive Pop III galaxies from photoionization feedback}
Motivated by recent observations, we examine how photoionization feedback could have led to a $\sim 10^7 ~M_\odot$ Pop III cluster in CR7. \cite{2015ApJ...808..139S} find that CR7 is consistent with clumps B and C hosting $\sim 10^{10}~M_\odot$ of older ``normal'' ($0.02~Z_\odot$) stars. As stars formed in these clumps at higher redshift, they ionized and photoheated their surroundings, potentially photoevaporating the gas in the progenitors of clump A. This may lead to a large Pop III starburst when clump A finally reaches the mass at which gas can overcome pressure and collapse into its halo at $z\sim6.6$.

\subsection{Clumps B and C}
We model clumps B and C as one halo (halo BC) with a mass of $M_{\rm BC} = 6.6\times10^{11}$ at $z=6.6$. We note that this mass is slightly lower than that assumed by \cite{2015arXiv151001733A} and \cite{2015arXiv151201111H}. Choosing a larger halo mass would not greatly impact the calculation discussed in this section, however it would reduce the predicted number density computed in \S 3 due to the sharp decline in the halo mass function at high mass \citep{1999MNRAS.308..119S}. 
The assumed halo mass implies a star formation efficiency of $f_* = 0.1$ to form $10^{10}~M_\odot$ of stars.

We model the history of ionized photon production from halo BC using a simple prescription based on dark matter halo merger trees. We construct merger trees with the method of \cite{2008MNRAS.383..557P} starting with $M_{\rm BC} = 6.6 \times 10^{11}~M_\odot$ at $z=6.6$ and using a mass resolution of $10^6 ~M_\odot$. Assuming that stars only form in atomic cooling halos, the star formation rate as a function of time is given by 
\begin{equation}
SFR_{\rm BC}(z) = f_{*}\frac{\Omega_{\rm b}}{\Omega_{\rm m}}\dot{M}_{\rm a,tot},
\end{equation}
where $f_*=0.1$ and $\dot{M}_{\rm a,tot}$ is the time derivative of the total mass in progenitors above the atomic cooling threshold. For the atomic cooling threshold, we assume a value of $M_{\rm a}=5.4\times 10^7 \left(\frac{1+z}{11} \right )^{-1.5}~M_\odot$, corresponding to the mass at which halos cool without molecular hydrogen in simulations \citep{2014MNRAS.439.3798F}.

\begin{figure}
\includegraphics[width=88mm]{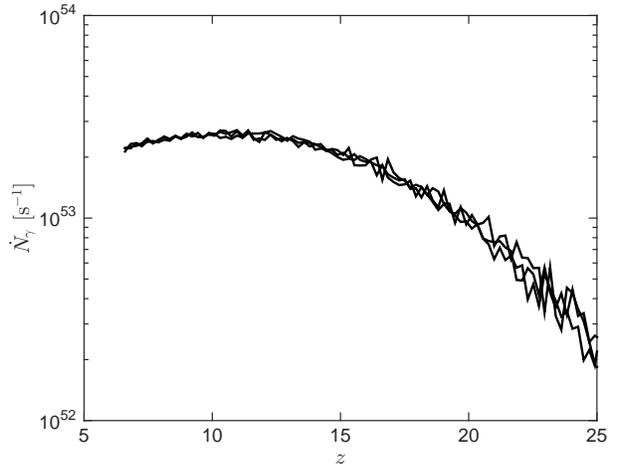}
\caption{\label{photon_plot} Ionizing photon luminosity from halo BC for three representative merger histories. We assume 4000 photons per baryon forming stars, a star formation efficiency of $f_* = 0.1$ in a $6.6 \times 10^{11}~M_\odot$ dark matter halo, and an escape fraction of $f_{\rm esc}=0.1$.}
\end{figure}

\begin{figure}
\includegraphics[width=88mm]{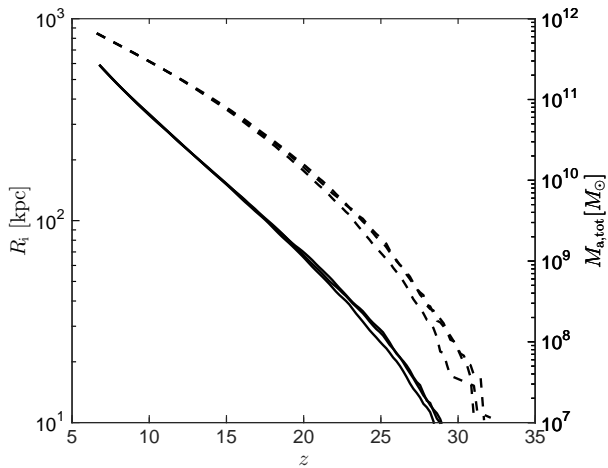}
\caption{\label{halo_BC_plot} Total mass in halos above the atomic cooling threshold in halo BC (right axis and dashed curves) and size of its ionized bubble (left axis and solid curves) for three representative merger histories.}
\end{figure}

To compute the ionizing flux and the size of the ionized bubble created by halo BC, we assume that 4000 ionizing photons are created for each baryon incorporated into stars, as produced by Pop II stars with a Salpeter IMF \citep[e.g. Table 1 in][]{2007MNRAS.377..285S}. This gives an ionized photon luminosity of
 $\dot{N}_\gamma = 4000 f_{\rm esc} \times SFR_{\rm BC}/m_{\rm p}$, where $m_{\rm p}$ denotes the proton mass and $f_{\rm esc}$ is the ionizing photon escape fraction. This quantity is plotted in Figure \ref{photon_plot}. We assume a fiducial value of $f_{\rm esc}=0.1$. The radius of the ionized bubble, $R_{\rm i}$, is then computed by solving
 \begin{equation}
\frac{dR^3_{\rm i}}{dt} = 3H(z)R^3_{\rm i} + \frac{3\dot{N}_\gamma}{4\pi \langle n_H \rangle } - C(z) \langle n_H \rangle \alpha_{\rm B} R^3_{\rm i},
\end{equation}
where $H(z)$ is the Hubble parameter, $\alpha_{\rm B}= 2.6\times 10^{-13} {\rm cm^3s^{-1}}$ is the case~B recombination coefficient of hydrogen at $T=10^4$ K, $\langle n_H \rangle$ is the mean cosmic hydrogen density, and $C(z) \equiv \langle n_{\rm HII}^2 \rangle / \langle n_{\rm HII} \rangle^2 $ is the clumping factor of the ionized IGM. We assume the clumping factor is given by 
\begin{equation}
C(z) =  2 \left ( \frac{1+z}{7}  \right )^{-2} + 1,
\end{equation}
which is similar to that found in simulations \citep{2015MNRAS.453.3593B, 2012MNRAS.427.2464F}. We also compute the LW luminosity from halo BC to determine the suppression of star formation in minihalos in the progenitors of clump A as discussed below. We assume that one LW photon is produced for each ionizing photon and an LW escape fraction of one. Note that for much smaller halos it has been shown that this escape fraction can be significantly lower \citep{2015MNRAS.454.2441S}.  In Figure \ref{halo_BC_plot}, we plot the ionized bubble radius and $M_{\rm a,tot}$ as a function of redshift for a few representative merger histories. It is clear that a large bubble is created at very early cosmic time.

\subsection{Clump A}
Next, we determine the likelihood that star formation in the progenitors of clump A was completely suppressed by LW radiation and photoevaporation from halo BC. We assume the mass of the halo hosting clump A (halo A) is approximately equal to the Jeans mass of photoevaporated gas near the virial radius of halo BC ($r_{\rm vir}\approx 40~{\rm kpc}$) at $z=6.6$. We choose this mass so that halo A could have just recently collapsed to form a Pop III starburst. The Jeans mass is given by \citep[e.g.][]{2014MNRAS.444..503N} 
\begin{multline}
\label{jeans_eqn}
M_{\rm J} = \frac{\pi^{5/2}}{6 \sqrt{\rho_{\rm m}}} \left (\frac{c_{\rm s}^2}{G} \right )^{3/2} \\ = 8\times 10^9 \left ( \frac{T}{10^4~{\rm K}} \right )^{3/2} \left( \frac{\rho_{\rm m}}{\bar{\rho}_{\rm m}(z=6.6)} \right )^{-1/2} M_\odot,
\end{multline}
where $c_{\rm s}$ is the sound speed, $T$ is the gas temperature, $\rho_{\rm m}$ is the total gas plus dark matter density, and $\bar{\rho}_{\rm m}(z)$ is the mean cosmic matter density. To reach the second equality we use the sound speed of ionized primordial gas with $\gamma = 5/3$. For an NFW profile with concentration of $c \sim 5$ \, the density at the virial radius is approximately 40 times the cosmic mean \citep{1997ApJ...490..493N} and the temperature of photoionized gas is approximately $10^4~{\rm K}$ \citep[see e.g. Figure 1 in ][]{2003ApJ...596....9H}. Plugging this into Eq.~\ref{jeans_eqn} gives $M_{\rm J} \approx 10^9~M_\odot$. Thus, we assume halo A has mass $M_{\rm A} = 10^9~M_\odot$ at $z=6.6$.

We utilize merger trees to determine the probability that no stars formed in halo A before $z\sim6.6$. Again we use the prescription of \cite{2008MNRAS.383..557P}, but in this case the mass resolution is set to $10^4~M_\odot$ to resolve minihalos. We assume that star formation has been suppressed if the progenitors of halo A are photoevaporated before the most massive progenitor halo (MMP) reaches the minimum mass where stars can form. We approximate this minimum mass with the smaller of the atomic cooling mass, $M_{\rm a}$, and  
\begin{equation}
M_{\rm m} =  2.5 \times 10^5 \left (  \frac{1+z}{26}  \right )^{-1.5} \left ( 1 + 6.96 \left (4\pi J_{\rm LW}(z) \right)^{0.47}\right ), 
\end{equation}
where $J_{\rm LW}$ is the LW background in units of $10^{-21}~{\rm erg~s^{-1}~cm^{-2}~Hz^{-1}~Sr^{-1}}$ \citep{2013MNRAS.432.2909F}. This formula gives a minimum mass consistent with the simulations of \citet{2001ApJ...548..509M},  \citet{2008ApJ...673...14O}, and  \citet{2007ApJ...671.1559W}. The mass for  $J_{\rm LW}=0$ is taken as the ``optimal fit'' from \cite{2012MNRAS.424.1335F} which was calibrated with the simulations of \citet{2011MNRAS.413..543S} and \citet{2011ApJ...736..147G}. We compute the LW flux from the growing BC halo by assuming one LW photon per ionizing photon produced as described in the previous subsection. We also include the contribution from the global LW background (taken from \cite{2015MNRAS.453.4456V}, in the case with metal enrichment, minihalo star formation efficiency of 0.001, and constant Pop II star formation efficiency), but find that the flux from halo BC dominates, except at the highest redshifts. We make the simplifying assumption that the distance between halo BC and halo A is constant and determine the fraction of realizations of halo A with suppressed star formation as a function of this separation.

To estimate the photoevaporation time we use the fitting formula of \cite{2005MNRAS.361..405I} (their Eq.~5 assuming a $5\times 10^4 ~ {\rm K}$ black body spectrum). Unfortunately, this formula only gives the time for a constant ionizing flux, while the flux from halo BC increases with time in our model (see Figure \ref{photon_plot}). Given this limitation, we make the conservative assumptions that the photoevaporation starts either at the redshift when the ionized bubble reaches halo A or $z=22$, whichever is lower, and that the photoevaporation time is given by the halo mass and ionizing flux at that redshift. We start photoevaporation at $z=22$ at the earliest because the flux from halo BC drops off rapidly at higher redshift. For our fiducial case, we assume that half the photoevaporation time given by \cite{2005MNRAS.361..405I} is sufficient to prevent all star formation. This corresponds to $\sim 90$ per cent of the gas being photoevaporated. We discuss how our results depend on this choice below.

\begin{figure}
\includegraphics[width=88mm]{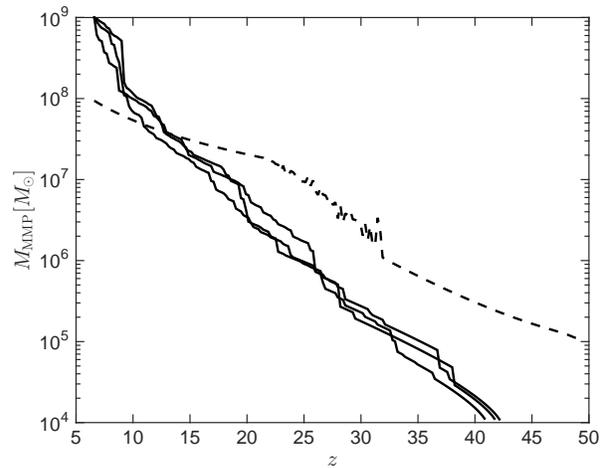}
\caption{ \label{halo_A_plot} The merger history for three realizations of halo A where star formation is completely suppressed before $z \sim 6.6$ (assuming a distance of 50 kpc from halo BC). The solid curves are the MMPs of halo A and the dashed curve is the minimum halo mass which results in star formation. The jagged portion of the minimum mass curve is due to the fluctuations in the LW radiation from halo BC, while at lower redshifts the curve corresponds to the atomic cooling mass and at higher redshifts the cosmic LW background dominates. The redshift where the MMP is photoevaporated is $z \sim 15$ for these merger histories (varying slightly depending on the mass of each halo when photoevaporation begins). }
\end{figure}

\begin{figure}
\includegraphics[width=88mm]{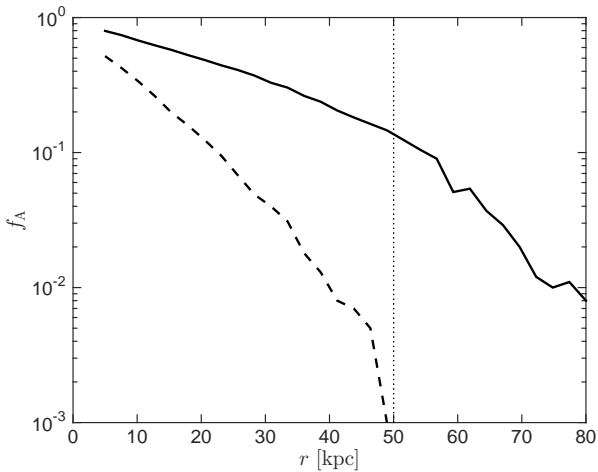}
\caption{\label{likelihood_plot} The probability that all star formation is suppressed in halo A as a function of the separation between halos A and BC. The solid curve is for the fiducial evaporation time and the dashed curve is for twice this time. The vertical dotted line corresponds to the approximate radius below which tidal forces are likely to prevent mass accretion on halo A.}
\end{figure}

With this information we are able to determine the probability that all star formation is completely suppressed in halo A. In Figure \ref{halo_A_plot}, we show examples of halo merger histories where this occurs. We test 1000 realizations and plot the likelihood that halo A does not form any stars by $z=6.6$ as a function of the separation between halos A and BC. This probability is plotted in Figure \ref{likelihood_plot}. Separations beyond $r \approx 50 ~ {\rm kpc}$ are most likely to produce a massive Pop III starburst. This is because at $r \lesssim r_{\rm vir, A} \left ( 2 M_{\rm BC}/M_{\rm A}  \right )^{1/3}$, where $r_{\rm vir, A}= 4.24~ {\rm kpc}$ is the virial radius of halo A, tidal forces from halo BC exceed the gravitational force binding halo A at the virial radius. This could prevent halo A from reaching $M_{\rm J}$. While the projected distance between the clumps in CR7 is $\sim5~{\rm kpc}$, they could be separated by a larger distance along the line of sight. In our fiducial case, with a separation roughly equal to 50 kpc we get a relatively high likelihood of $\sim 0.14$. Note that we only use one merger history for halo BC. Since we are considering the total mass above the atomic cooling threshold there is less variability than the MMP and thus changing the realization of BC does not have a large impact on our results.

\section{Abundance of Pop III Clusters}
With an estimate of the fraction of halos similar to halo A which potentially produce a large Pop III starburst at $z\sim6.6$, we compute a rough approximation of the global abundance of similar objects which should be visible at this redshift. This abundance is given by
\begin{equation}
n_{\rm PopIII} = n_{\rm BC} \frac{dN_{\rm A}}{dt} t_{\rm duty} f_{\rm A},
\end{equation}
where $n_{\rm BC}$ is the density of dark matter halos above $M_{\rm BC}$ (most of which are very near $M_{\rm BC}$), $\frac{dN_{\rm A}}{dt}$ is the rate at which $M_{\rm A}$ sized halos merge with $M_{\rm BC}$ sized halos, $t_{\rm duty}$ is the time that the Pop III starburst is visible, and $f_{\rm A}$ is the fraction of halo A realizations that stay pristine due to complete suppression of star formation computed above. As we discuss above, we believe the most relevant distance to evaluate $f_{\rm A}$ at is 50 kpc, due to tidal disruption at small separation. We assume $t_{\rm duty}\approx 3~{\rm Myr}$ 
corresponding roughly to the age of the stars inferred by \cite{2015ApJ...808..139S}. An order of magnitude estimation of $\frac{dN_{\rm A}}{dt}$ is obtained by examining the merger trees described above. On average, $\sim150$ Myr before $z=6.6$ the MMP of halo A is $6.3 \times 10^8~M_\odot$. At this time, a realization of halo BC has $\sim 8$ progenitor halos between this mass and $10^9~M_\odot$. Thus, roughly 8 halos similar to halo A merge with halo BC each 150 Myr, giving $\frac{dN_{\rm A}}{dt} \approx 0.05~ {\rm Myr}^{-1}$. Putting this all together we get an abundance of massive Pop III starbursts of $\approx 10^{-7}~{\rm Mpc^{-3}}$. This is roughly one order of magnitude less than the density of observed Ly$\alpha$ emitters as bright as CR7 \citep[computed with the power law luminosity function of][]{2015MNRAS.451..400M}. While we stress that our estimate is highly uncertain, we conclude that our scenario could potentially explain the CR7 observations (since not all Ly$\alpha$ emitters that bright need to contain Pop III stars). Even if it does not, it is quite possible that a massive Pop III galaxy similar to what we predict could be observed in a larger survey.

\section{Discussion and Conclusions}
We have explored a scenario for the formation of massive Pop III star clusters based on photoionization feedback. Motivated by the observations of CR7, we consider a bright source (halo BC) which completely suppresses star formation in the progenitors of a $\sim 10^9~M_\odot$ dark matter halo (halo A) through LW feedback and photoevaporation. Once this halo reaches the Jeans mass ($M_{\rm J}\sim 10^9~M_\odot$), gas is able to collapse into the halo, potentially producing a massive burst of Pop III stars. We utilized merger trees to determine the probability that a $\sim 10^9~M_\odot$ halo forming near a $6.6 \times 10^{11}~M_\odot$ halo will have no star formation before $z=6.6$ as a function of separation. We find that for a separation of $50~{\rm kpc}$ this probability is $\approx 0.14$. Using this probability, we estimate that the number density of similar objects visible at $z=6.6$ is $\approx 10^{-7}~{\rm Mpc^{-3}}$.

Here we discuss the assumptions and approximations that went into the above calculations. First, we assume that once halo A reaches $M_{\rm J} \approx 10^9~M_\odot$ it quickly forms $\sim10^7 ~ M_\odot$ of Pop III stars. For the stars to form within the lifetime of large Pop III stars ($\sim$a few Myr), before metals are injected into the gas, the star formation rate needs to be greater than a few $M_\odot/{\rm yr}$ and continue until $\sim 7$ per cent of the halo gas is incorporated into stars. As a rough indication, the gas mass divided by the dynamical time at the virial radius of halo A is approximately equal to this required rate. The rate of infall towards the center of the halo could be significantly higher. The required star formation efficiency is similar to that inferred in high redshift galaxies through abundance matching \citep[see Figure 3 in][]{2015MNRAS.453.4456V}. This suggests that it is plausible for such a large mass of Pop III stars to form, however this is highly uncertain and the required mass function needs to be checked in future hydrodynamical simulations. We also note that the mass at which gas collapses into the halo could be larger than our assumed value of $\sim 10^9~M_\odot$, reducing the required star formation efficiency. Additionally, at a separation of $50~{\rm kpc}$, the LW flux on halo A is $\sim 20 \times 10^{-21} {\rm erg~s^{-1}~Hz{-1}~cm^{-2}~sr^{-1}}$ at $z=6.6$. This is lower than required to produce a DCBH in an atomic cooling halo \citep[][]{2010MNRAS.402.1249S, 2014MNRAS.445..544S}. In simulations of recently collapsed atomic cooling halos (which are substantially smaller than our halo A), the gas is sufficiently dense to produce the HeII and Ly$\alpha$ recombination line emission observed in CR7 \citep{2014MNRAS.439.1160R}. These same halos can have turbulent velocities a factor of a few larger than their circular velocities, which may explain the $\sim 100 {\rm km~s^{-1}}$ HeII 1640~\AA~line width observed in CR7.

Another simplifying assumption we have made is that the gas in halo A is not metal enriched by supernovae winds from BC. For a separation of $\sim 50~{\rm kpc}$, it would take a $50 ~ {\rm km~s^{-1}}$ wind $1~{\rm Gyr}$ to travel the distance between halos BC and A. If this wind is launched at $z=22$ (the redshift where the SFR in halo BC increases and when we assume photoevaporation begins), it would not reach halo A by $z=6.6$ (this difference in redshift corresponds to 700 Myr). Even if the winds are faster, it is also possible that they could spread metals asymmetrically, preventing them from reaching halo A in time to prevent Pop III star formation.

We assumed a characteristic separation between halos A and BC of $\sim 50~{\rm kpc}$. This distance corresponds to the separation where the tidal forces on halo A equal the self-gravitational forces binding halo A together (evaluated at the virial radius of halo A). At smaller separations we calculate a higher probability of photoevaporation, however tidal interactions would most likely prevent the growth of halo A and it would never reach the Jeans mass. We note that while we assume a constant separation, the actual physical separation would vary as a function of time. To get a rough understanding of this time variation, we consider the trajectory of the outermost shell of halo BC using the spherical collapse model. This corresponds roughly to the position of halo A. At $z=22$, when photoevaporation begins, the separation is $\sim 55~{\rm kpc}$. The maximum distance of $\sim 70 ~{\rm kpc}$ is reached at $z\sim11$, and the distance rapidly decreases (to zero in the spherical collapse model) by $z=6.6$. The larger distance before collapse is likely to help prevent tidal forces from stopping the final accretion of dark matter before the halo reaches $M_{\rm J}$. The increased distance could reduce the flux on halo A by a factor of $\sim2$ as it completes photoevaporation, however this is comparable to or smaller than other uncertainties in our calculation (e.g. the escape fraction of ionizing photons).

When computing the star formation history of halo BC, we assume that the star formation rate is proportional to the growth of the halo progenitors, which gives a value as a function of time proportional to the curve in Figure \ref{photon_plot}. In contrast to our simple merger tree model, clumps B and C in CR7 are not currently undergoing active star formation. This may help to form a massive Pop III galaxy, since the ionizing radiation could have been stronger early on, more quickly photoevaporating the progenitor halos of clump A.

The photoevaporation time is one of the most important and uncertain quantities in the calculation presented above. We are conservative in our assumption that the flux at $z=22$ is the one used to compute this time even though the flux is still increasing. However, for our fiducial case we assume that half the photoevaporation time given by the fits of \cite{2005MNRAS.361..405I} is sufficient to prevent all star formation in the MMP of halo A. This correspsonds to $\sim 90$ per cent of the gas being photoevaporated in \cite{2005MNRAS.361..405I}. Our results are sensitive to the exact value of this assumed timescale. Increasing the photoevaporation timescale by a factor of two reduces the abundance of massive Pop III stars by more than an order of magnitude.

In conclusion, we consider the formation of massive Pop III galaxies through photoionization feedback to be a plausible explanation for CR7.
We estimate that the number density of objects formed this way is $\approx 10^{-7}~ {\rm Mpc^{-3}}$ (but emphasize this is highly uncertain). This is ten times lower than the observed density of Ly$\alpha$ emitters as bright as CR7 \citep{2015MNRAS.451..400M} (which suggests that only a fraction of these bright sources will harbor Pop III stars). Without the suppression of star formation by photoionization discussed here, the dark matter halo hosting clump A in CR7 is likely to have formed stars once it reached the atomic cooling threshold ($\sim 10^8~M_\odot$ at $z=6.6$). Even assuming that none of the gas in clump A has been photoevaporated, 50 per cent of this gas would need to be converted into Pop III stars before it is self-polluted by supernovae winds to reach the $10^7~M_\odot$ implied by CR7. Such a high star formation efficiency seems unlikely and thus we favor the scenario described here. While we find this channel of Pop III star formation to be promising, future detailed hydrodynamical simulations are required to test its viability. Finally, we point out that even if CR7 did not form through the photoevaporation feedback mechanism, there may be other massive Pop III galaxies formed this way which could be observed in large Ly$\alpha$ emitter surveys and quickly confirmed with future observations from the James Webb Space Telescope.

\section*{Acknowledgements}
EV was supported by the Columbia Prize Postdoctoral Fellowship in the Natural Sciences.  ZH was supported by NASA grant NNX15AB19G.  GLB was supported by National Science Foundation grants 1008134 and ACI-1339624 and NASA grant NNX15AB20G.

\bibliography{paper}

\begin{thebibliography}{}

\bibitem[\protect\citeauthoryear{{Ade}, {Aghanim}, {Armitage-Caplan}, {Arnaud},
  {Ashdown} \& et al.}{{Ade} et~al.}{2014}]{2014A&A...571A..16P}
{Ade} P.~A.~R.,  {Aghanim} N.,  {Armitage-Caplan} C.,  {Arnaud} M.,  {Ashdown}
  M.,    et al. 2014, \aap, 571, A16

\bibitem[\protect\citeauthoryear{{Agarwal}, {Johnson}, {Zackrisson}, {Labbe},
  {van den Bosch}, {Natarajan} \& {Khochfar}}{{Agarwal}
  et~al.}{2015}]{2015arXiv151001733A}
{Agarwal} B.,  {Johnson} J.~L.,  {Zackrisson} E.,  {Labbe} I.,  {van den Bosch}
  F.~C.,  {Natarajan} P.,    {Khochfar} S.,  2015, ArXiv e-prints: 1510.01733

\bibitem[\protect\citeauthoryear{{Bauer}, {Springel}, {Vogelsberger}, {Genel},
  {Torrey}, {Sijacki}, {Nelson} \& {Hernquist}}{{Bauer}
  et~al.}{2015}]{2015MNRAS.453.3593B}
{Bauer} A.,  {Springel} V.,  {Vogelsberger} M.,  {Genel} S.,  {Torrey} P.,
  {Sijacki} D.,  {Nelson} D.,    {Hernquist} L.,  2015, \mnras, 453, 3593

\bibitem[\protect\citeauthoryear{{Fernandez}, {Bryan}, {Haiman} \&
  {Li}}{{Fernandez} et~al.}{2014}]{2014MNRAS.439.3798F}
{Fernandez} R.,  {Bryan} G.~L.,  {Haiman} Z.,    {Li} M.,  2014, \mnras, 439,
  3798

\bibitem[\protect\citeauthoryear{{Fialkov}, {Barkana}, {Tseliakhovich} \&
  {Hirata}}{{Fialkov} et~al.}{2012}]{2012MNRAS.424.1335F}
{Fialkov} A.,  {Barkana} R.,  {Tseliakhovich} D.,    {Hirata} C.~M.,  2012,
  \mnras, 424, 1335

\bibitem[\protect\citeauthoryear{{Fialkov}, {Barkana}, {Visbal},
  {Tseliakhovich} \& {Hirata}}{{Fialkov} et~al.}{2013}]{2013MNRAS.432.2909F}
{Fialkov} A.,  {Barkana} R.,  {Visbal} E.,  {Tseliakhovich} D.,    {Hirata}
  C.~M.,  2013, \mnras, 432, 2909

\bibitem[\protect\citeauthoryear{{Finlator}, {Oh}, {{\"O}zel} \&
  {Dav{\'e}}}{{Finlator} et~al.}{2012}]{2012MNRAS.427.2464F}
{Finlator} K.,  {Oh} S.~P.,  {{\"O}zel} F.,    {Dav{\'e}} R.,  2012, \mnras,
  427, 2464

\bibitem[\protect\citeauthoryear{{Greif}}{{Greif}}{2015}]{2015ComAC...2....3G}
{Greif} T.~H.,  2015, Computational Astrophysics and Cosmology, 2, 3

\bibitem[\protect\citeauthoryear{{Greif}, {White}, {Klessen} \&
  {Springel}}{{Greif} et~al.}{2011}]{2011ApJ...736..147G}
{Greif} T.~H.,  {White} S.~D.~M.,  {Klessen} R.~S.,    {Springel} V.,  2011,
  \apj, 736, 147

\bibitem[\protect\citeauthoryear{{Haiman}, {Rees} \& {Loeb}}{{Haiman}
  et~al.}{1997}]{1997ApJ...476..458H}
{Haiman} Z.,  {Rees} M.~J.,    {Loeb} A.,  1997, \apj, 476, 458

\bibitem[\protect\citeauthoryear{{Hartwig}, {Latif}, {Magg}, {Bromm},
  {Klessen}, {Glover}, {Whalen}, {Pellegrini} \& {Volonteri}}{{Hartwig}
  et~al.}{2015}]{2015arXiv151201111H}
{Hartwig} T.,  {Latif} M.~A.,  {Magg} M.,  {Bromm} V.,  {Klessen} R.~S.,
  {Glover} S.~C.~O.,  {Whalen} D.~J.,  {Pellegrini} E.~W.,    {Volonteri} M.,
  2015, ArXiv e-prints: 1512.01111

\bibitem[\protect\citeauthoryear{{Hui} \& {Haiman}}{{Hui} \&
  {Haiman}}{2003}]{2003ApJ...596....9H}
{Hui} L.,  {Haiman} Z.,  2003, \apj, 596, 9

\bibitem[\protect\citeauthoryear{{Iliev}, {Shapiro} \& {Raga}}{{Iliev}
  et~al.}{2005}]{2005MNRAS.361..405I}
{Iliev} I.~T.,  {Shapiro} P.~R.,    {Raga} A.~C.,  2005, \mnras, 361, 405

\bibitem[\protect\citeauthoryear{{Johnson}}{{Johnson}}{2010}]{2010MNRAS.404.1425J}
{Johnson} J.~L.,  2010, \mnras, 404, 1425

\bibitem[\protect\citeauthoryear{{Machacek}, {Bryan} \& {Abel}}{{Machacek}
  et~al.}{2001}]{2001ApJ...548..509M}
{Machacek} M.~E.,  {Bryan} G.~L.,    {Abel} T.,  2001, \apj, 548, 509

\bibitem[\protect\citeauthoryear{{Matthee}, {Sobral}, {Santos},
  {R{\"o}ttgering}, {Darvish} \& {Mobasher}}{{Matthee}
  et~al.}{2015}]{2015MNRAS.451..400M}
{Matthee} J.,  {Sobral} D.,  {Santos} S.,  {R{\"o}ttgering} H.,  {Darvish} B.,
    {Mobasher} B.,  2015, \mnras, 451, 400

\bibitem[\protect\citeauthoryear{{Navarro}, {Frenk} \& {White}}{{Navarro}
  et~al.}{1997}]{1997ApJ...490..493N}
{Navarro} J.~F.,  {Frenk} C.~S.,    {White} S.~D.~M.,  1997, \apj, 490, 493

\bibitem[\protect\citeauthoryear{{Noh} \& {McQuinn}}{{Noh} \&
  {McQuinn}}{2014}]{2014MNRAS.444..503N}
{Noh} Y.,  {McQuinn} M.,  2014, \mnras, 444, 503

\bibitem[\protect\citeauthoryear{{Oh}, {Haiman} \& {Rees}}{{Oh}
  et~al.}{2001}]{2001ApJ...553...73O}
{Oh} S.~P.,  {Haiman} Z.,    {Rees} M.~J.,  2001, \apj, 553, 73

\bibitem[\protect\citeauthoryear{{O'Shea} \& {Norman}}{{O'Shea} \&
  {Norman}}{2008}]{2008ApJ...673...14O}
{O'Shea} B.~W.,  {Norman} M.~L.,  2008, \apj, 673, 14

\bibitem[\protect\citeauthoryear{{Pallottini}, {Ferrara}, {Pacucci},
  {Gallerani}, {Salvadori}, {Schneider}, {Schaerer}, {Sobral} \&
  {Matthee}}{{Pallottini} et~al.}{2015}]{2015MNRAS.453.2465P}
{Pallottini} A.,  {Ferrara} A.,  {Pacucci} F.,  {Gallerani} S.,  {Salvadori}
  S.,  {Schneider} R.,  {Schaerer} D.,  {Sobral} D.,    {Matthee} J.,  2015,
  \mnras, 453, 2465

\bibitem[\protect\citeauthoryear{{Parkinson}, {Cole} \& {Helly}}{{Parkinson}
  et~al.}{2008}]{2008MNRAS.383..557P}
{Parkinson} H.,  {Cole} S.,    {Helly} J.,  2008, \mnras, 383, 557

\bibitem[\protect\citeauthoryear{{Regan}, {Johansson} \& {Haehnelt}}{{Regan}
  et~al.}{2014}]{2014MNRAS.439.1160R}
{Regan} J.~A.,  {Johansson} P.~H.,    {Haehnelt} M.~G.,  2014, \mnras, 439,
  1160

\bibitem[\protect\citeauthoryear{{Samui}, {Srianand} \& {Subramanian}}{{Samui}
  et~al.}{2007}]{2007MNRAS.377..285S}
{Samui} S.,  {Srianand} R.,    {Subramanian} K.,  2007, \mnras, 377, 285

\bibitem[\protect\citeauthoryear{{Schauer}, {Whalen}, {Glover} \&
  {Klessen}}{{Schauer} et~al.}{2015}]{2015MNRAS.454.2441S}
{Schauer} A.~T.~P.,  {Whalen} D.~J.,  {Glover} S.~C.~O.,    {Klessen} R.~S.,
  2015, \mnras, 454, 2441

\bibitem[\protect\citeauthoryear{{Shang}, {Bryan} \& {Haiman}}{{Shang}
  et~al.}{2010}]{2010MNRAS.402.1249S}
{Shang} C.,  {Bryan} G.~L.,    {Haiman} Z.,  2010, \mnras, 402, 1249

\bibitem[\protect\citeauthoryear{{Sheth} \& {Tormen}}{{Sheth} \&
  {Tormen}}{1999}]{1999MNRAS.308..119S}
{Sheth} R.~K.,  {Tormen} G.,  1999, \mnras, 308, 119

\bibitem[\protect\citeauthoryear{{Sobral}, {Matthee}, {Darvish}, {Schaerer},
  {Mobasher}, {R{\"o}ttgering}, {Santos} \& {Hemmati}}{{Sobral}
  et~al.}{2015}]{2015ApJ...808..139S}
{Sobral} D.,  {Matthee} J.,  {Darvish} B.,  {Schaerer} D.,  {Mobasher} B.,
  {R{\"o}ttgering} H.~J.~A.,  {Santos} S.,    {Hemmati} S.,  2015, \apj, 808,
  139

\bibitem[\protect\citeauthoryear{{Stacy}, {Bromm} \& {Loeb}}{{Stacy}
  et~al.}{2011}]{2011MNRAS.413..543S}
{Stacy} A.,  {Bromm} V.,    {Loeb} A.,  2011, \mnras, 413, 543

\bibitem[\protect\citeauthoryear{{Sugimura}, {Omukai} \& {Inoue}}{{Sugimura}
  et~al.}{2014}]{2014MNRAS.445..544S}
{Sugimura} K.,  {Omukai} K.,    {Inoue} A.~K.,  2014, \mnras, 445, 544

\bibitem[\protect\citeauthoryear{{Tumlinson}, {Giroux} \& {Shull}}{{Tumlinson}
  et~al.}{2001}]{2001ApJ...550L...1T}
{Tumlinson} J.,  {Giroux} M.~L.,    {Shull} J.~M.,  2001, \apjl, 550, L1

\bibitem[\protect\citeauthoryear{{Visbal}, {Haiman} \& {Bryan}}{{Visbal}
  et~al.}{2015}]{2015MNRAS.453.4456V}
{Visbal} E.,  {Haiman} Z.,    {Bryan} G.~L.,  2015, \mnras, 453, 4456

\bibitem[\protect\citeauthoryear{{Visbal}, {Haiman}, {Terrazas}, {Bryan} \&
  {Barkana}}{{Visbal} et~al.}{2014}]{2014MNRAS.445..107V}
{Visbal} E.,  {Haiman} Z.,  {Terrazas} B.,  {Bryan} G.~L.~.,    {Barkana} R.,
  2014, \mnras, 445, 107

\bibitem[\protect\citeauthoryear{{Wise} \& {Abel}}{{Wise} \&
  {Abel}}{2007}]{2007ApJ...671.1559W}
{Wise} J.~H.,  {Abel} T.,  2007, \apj, 671, 1559

\bibitem[\protect\citeauthoryear{{Wise}, {Turk}, {Norman} \& {Abel}}{{Wise}
  et~al.}{2012}]{2012ApJ...745...50W}
{Wise} J.~H.,  {Turk} M.~J.,  {Norman} M.~L.,    {Abel} T.,  2012, \apj, 745,
  50

\bibitem[\protect\citeauthoryear{{Wolcott-Green}, {Haiman} \&
  {Bryan}}{{Wolcott-Green} et~al.}{2011}]{2011MNRAS.418..838W}
{Wolcott-Green} J.,  {Haiman} Z.,    {Bryan} G.~L.,  2011, \mnras, 418, 838

\end{thebibliography}

\end{document}